\documentclass[runningheads]{llncs}
\usepackage{graphicx}

\usepackage{amsmath}
\usepackage[colorinlistoftodos]{todonotes}
\usepackage[colorlinks=true, allcolors=blue]{hyperref}

\begin{document}
\title{Building an Ecosystem for the Tyrolean Tourism Knowledge Graph}
\author{Elias K\"arle \orcidID{0000-0002-2686-3221} \and
Umutcan \c{S}im\c{s}ek \orcidID{0000-0001-6459-474X} \and
Oleksandra Panasiuk \and
Dieter Fensel
}
\authorrunning{E. K\"arle et al.}

\institute{Semantic Technology Institute, Universit\"at Innsbruck,\\Technikerstrasse 21a, 6020 Innsbruck, Austria\\
\email{\{firstname.lastname\}@sti2.at}}
\maketitle 
\begin{abstract}
The introduction of the schema.org vocabulary was a big step towards making websites machine read- and understandable. Due to schema.org's RDF-like nature storing annotations in a graph database is easy and efficient. In this paper the authors show how they gather touristic data in the Austrian region of Tirol and provide this data publicly in a knowledge graph. The definition of subsets of the vocabulary is followed by providing means to map data sources efficiently to schema.org and then store the annotated content into the graph. To showcase the consumption of the touristic data four scenarios are described which use the knowledge graph for real life applications and data analysis.

\end{abstract}

\section{Introduction}
\label{sec:Introduction}

The term knowledge graph has been coined by Google in 2012\footnote{\url{https://googleblog.blogspot.co.at/2012/05/introducing-knowledge-graph-things-not.html}}. There is no formal definition for the term, however based on several definition attempts we consider knowledge graphs as "a knowledge base in graph form that mostly contains real-world entities and their relationships".
 Having the "knowledge" in graph form presents many advantages over relational data storage such as scalable growth, due to the lack of data and schema distinction. From an implementation point of view, standards and tools from the semantic web stack (e.g. RDF, SPARQL, Triple Stores) can be utilized, which accelerates the knowledge graph construction and consumption by automated agents like Intelligent Personal Assistants (IPAs).
Tourism is a major economic sector in Tirol, generating around 20\% of GDP in the region\footnote{\url{https://presse.tirol.at/de/daten-zahlen-zum-tourismus-in-tirol/pr335467}}. The Tyrolean tourism sector has been going through a extensive digitalization movement in the recent years. This movement includes making the valuable data hidden behind the walls of proprietary systems available on the web for automated agents like dialogue systems and intelligent personal assistants. Given the aforementioned factors, the most scalable way to achieve this is to enrich the data semantically and publish it on the web to facilitate the creation of a knowledge graph.

In this paper, we present the ecosystem we have been building for the creation of a Tyrolean Knowledge Graph that contributes to the digitalization of tourism in the region. We explain how we help various parties in the tourism sector publishing semantically annotated data with schema.org\footnote{\url{https://schema.org/}} and how we transfer the collected data into a knowledge graph. Additionally, we demonstrate various consumption scenarios for our knowledge graph.

The remainder of the paper is structured as follows: Section \ref{sec:RelatedWork} gives an overview to the existing knowledge graphs and explains our motivation. Section \ref{sec:Graph} shows how we construct the knowledge graph and awaiting challenges. Section \ref{sec:UseCase} presents various scenarios where the knowledge graph can be utilized. Section \ref{sec:Conclusion} concludes the paper with a summary and an outlook.

\section{Related Work and Motivation}
\label{sec:RelatedWork}
Following the semantic web \cite{berners2001semantic} and linked data \cite{bizer_linked_2009} research, many open knowledge graphs have been published. The initial prominent efforts aimed to cover as many domains as possible, therefore used Wikipedia as a source to construct the knowledge graphs. The DBPedia and YAGO knowledge graphs uses the infoboxes and categories in the Wikipedia website to extract triples. While Wikidata \cite{vrandecic2014} follows a more collaborative approach and benefit from both bot and human contributions and focus on the provenance of the data. The NELL \cite{zimmermann2013nell2rdf} project aims to extract triples from unstructured text by crawling the web.
As for the tourism domain, the closest work to ours is the 3cixty \cite{TRONCY20172} project. The project aims to create knowledge graphs for smart city applications. They use an ontology that heavily reuses schema.org to describe their data.

The Tyrolean Tourism Knowledge Graph contains static (e.g. phone number, address) and dynamic (e.g. accommodation offers) data based on schema.org annotations collected from different sources such as Destination Management Organizations (DMO) and Geographical Information Systems (GIS). In our previous work \cite{akbar2017complete}, we explained how we annotated the relevant data in the region with schema.org from different sources. Our knowledge graph consolidates these annotations and enables intelligent applications like chatbots to contribute the digitalization of tourism in Tyrol. Additionally, since we store the historical data, the knowledge graph allows data analytics to provide insights from the region.

\section{Feeding the Graph}
\label{sec:Graph}
The quality of a knowledge graph is highly dependent on the data it contains. To build a sustainable, high quality tourism knowledge graph, reliable data sources have to be identified. The best source of frequently updated touristic data is of course a hotel website. But an analysis of the distribution of schema.org amongst hotels\cite{karle2016there} showed, that the current state of schema.org in tourism and especially in the accommodation business has too little adoption to be used as a data source. Yet, due to its growing uptake, driven by the big search engine providers, we still decided to work with schema.org, and to first extend the schema.org vocabulary for the accommodation sector\cite{kaerle2017extending} (released as part of schema.org 3.1\footnote{\url{http://schema.org/docs/releases.html\#v3.1}}) and then fostering the distribution of schema.org in the whole tourism sector. Only having touristic websites annotated with schema.org would ensure the repeatability of the data aggregation process in the long run and at the same time help the touristic websites make their content more visible by implicitly applying semantic search engine optimization.

To simplify and unify the annotation process we started by defining sets of vocabularies for specific domains. The idea of the resulting domain specifications (DS) was first published in \cite{Simsek2017} and later applied to tourism in \cite{panasiuk2017defining}. The DS are subsets of the schema.org vocabulary, each associated with required and optional properties. They provide the recommended patterns for annotating different touristic domains such as hotels, restaurants, ski schools and others and define the model of the structured data on the web. The domain specification can be consider as schema.org design patterns.

Instead of annotating the actual accommodation websites one by one we approached different Tyrolean destination management organizations (DMOs) and their IT service providers. For accommodation data, data about the regional events and infrastructure we worked with Feratel\footnote{\url{http://www.feratel.com/}}, a full stack touristic IT service provider, and Infomax\footnote{\url{https://www.infomax.de}}. For geodata we cooperated with General Solutions, a company specialized on visualizing geospatial information on web maps\footnote{\url{https://general-solutions.eu/}}. A generic source for touristic data we annotated was Outdooractive\footnote{\url{https://www.outdooractive.com/}} and besides that also data about ski schools and ski lessons, provided by the company Waldhart Software\footnote{\url{https://www.waldhart.at/}}. Information not yet covered by the mentioned data sources was collected and annotated from the DMO's website directly.

For the annotation of web content we had to make a distinction between three different types of data. Static data is information about the core data of a business, like its address or a description. Dynamic data describes things like availabilities and prices and frequently changes. Active data describes software interfaces to interact with, like for example a booking API. The DS are sufficient for the manual annotation of static data. For dynamic data this is not an option. We decided to build wrapper software to allow automatic annotation. A wrapper defines different mappings from a data source to schema.org. Then the wrapper software reads the source, maps the data and stores the resulting file. To see the DS and the wrapper software in action we implemented these features in semantify.it, a SaaS platform for creation, validation and distribution of semantic annotations. This platform stores the annotations as individual JSON-LD files in a MongoDB collection for more convenient publication on web pages. A detailed description of the semantify.it platform was published in \cite{karle2017semantify}. Data sources we applied those wrappers to are, as mentioned above, Feratel, General Solutions, Infomax and Outdooractive. The wrapper made for General Solutions was also published in \cite{panasiuk18geodata}.

With that tools at hand we could start rolling out annotations into tourism. To have a maximum impact we approached different destination management organizations (DMOs) and applied our annotation process to their comprehensive websites. The results was the complete annotation of a DMO's static data and can be found in \cite{akbar2017complete}. To ensure the quality of the annotations we also enhanced the DS to apply rules to touristic domains. Trough that extension a validator described in \cite{Simsek2017} can not only check the syntactic, but also the semantic correctness of annotations.

While with the mentioned solutions the annotation of static data was straight forward, the annotation of dynamic data raised some problems. Accommodation businesses, for example, offer several rooms, with different pricing- and catering options with different occupancy possibilities, being charge at changing rates at different seasons over the year, and on top with flexible stay durations. The result is a vast amount of booking options which goes, if expressed in schema.org annotations, into annotation file sizes of megabytes. To avoid this materialization of booking possibilities we developed the idea of publication heuristics to enable a representation of offers in schema.org for a website\cite{kaerle2018heuristics}.

Finally, we needed to annotate the active data to allow bookings and purchases of touristic products and services by automated agents or third party software applications. Therefore we developed a way to annotated web APIs with the schema.org vocabulary and hence represent them as lightweight semantic web services\cite{simsek2018apiwrapping}. The resulting "action wrapper" was applied to the internet booking engine software providers Easybooking, Feratel and Kognitiv. Both, the publication heuristics and the action wrapper were implemented as parts of the semantify.it platform.

To fill the knowledge graph with the curated data we replicated the data stored in semantify.it into the graph and added crawled schema.org annotations of already annotated touristic websites on the fly. For the replication process we tested two approaches. One used Java, taking the intermediate step of translating the data from JSON-LD to RDF and only then writing it into the graph. The other approach, which proofed to be the more efficient one, writes JSON-LD directly into the graph. Both approaches were using basic cleaning and identification measures on the data, where for example every website's crawl was stored in an explicit named graph for later reuse of legacy data for reasoning purposes. Later, more advanced data polishing measures will include the removal of duplicates, consolidation of entities and enrichment of data sets by adding information form trusted other sources.

We perform daily data migrations from semantify.it to the knowledge graph since December 2017. To build the knowledge graph, we use GraphDB\footnote{\url{http://graphdb.ontotext.com/}}, a product by Ontotext. Our knowledge graph currently contains 1.5 Billion statements of which 800 Million are explicit and 700 Million are inferred\footnote{April 2018 numbers}.

\begin{figure}
\includegraphics[width=\linewidth]{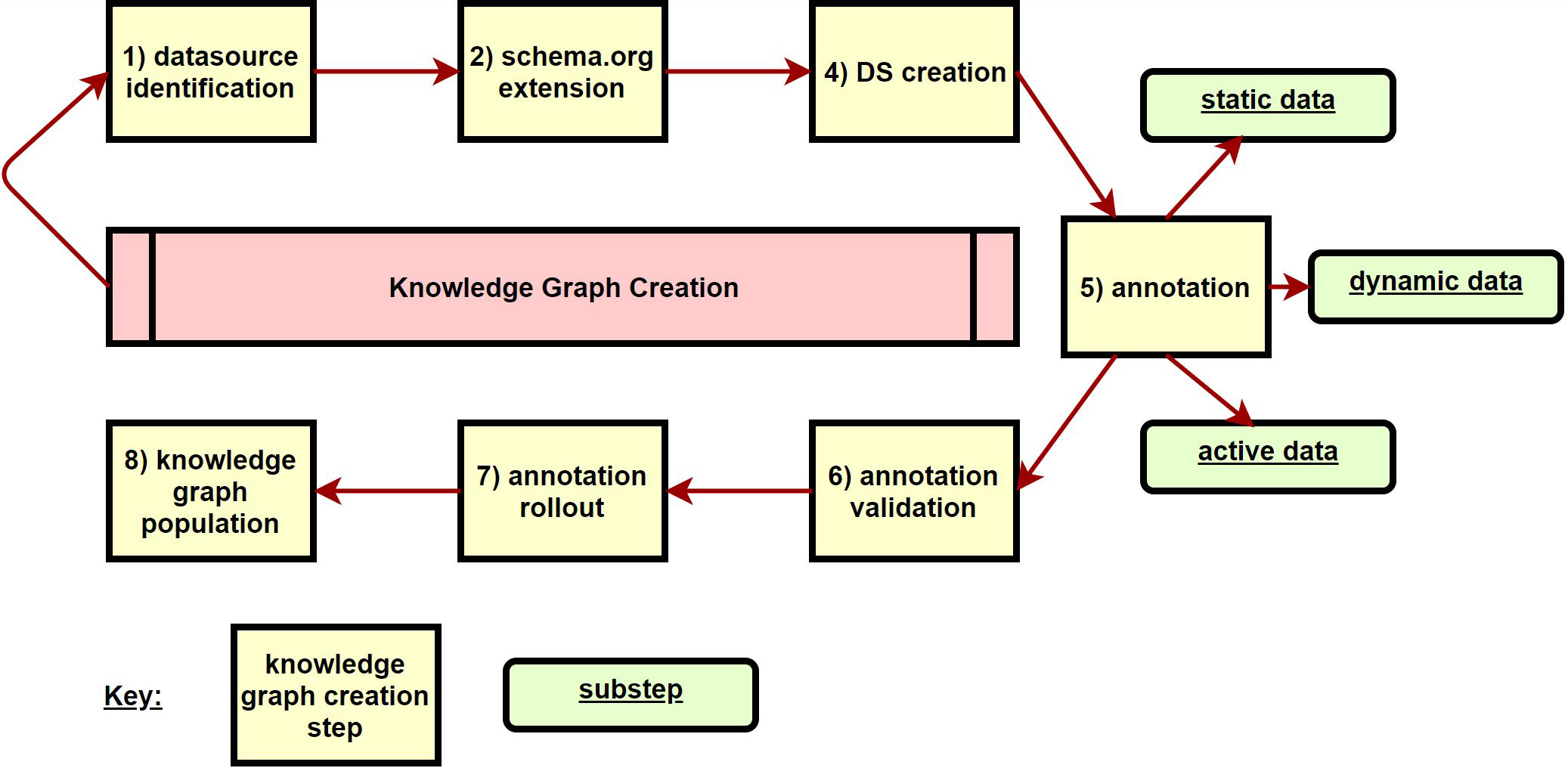}
\caption{The eight steps to create the Tirolean Tourism Knowledge Graph.}
\label{fig:lifecycle}
\end{figure}

An overview of the knowledge graph creation life cycle according to our survey can be found in Figure \ref{fig:lifecycle}.

\section{Use Case}
\label{sec:UseCase}
The touristic data stored in the knowledge graph is reachable via a SPARQL interface\footnote{\url{http://graphdb.sti2.at:8080/sparql}} and open to everyone. For the purpose of demonstration and experimentation we applied four different use case scenarios on the graph's data.

\subsection{Conversational Assistants}
To demonstrate how the annotations, stored in knowledge graph, can be interpreted in a way accessible for users, we developed the conversation tourism assistant \cite{panasiuk2018eswc}. This agent extracts the core information items from the user's questions and maps them to schema.org types and properties. As dialog interface we used Google's Dialogflow\footnote{\url{https://dialogflow.com}} and develop the web service to discover requested information from the knowledge graph through SPARQL queries. The result is a fully functional chatbot-like assistant system on top of our GraphDB database. Deployed to Amazon's Alexa for example it is possible to ask things like "Alexa, I want to do hiking in the region around Seefeld". The request is understood by Dialogflow and then forwarded to our web hook which does the mapping and the data retrieval from the graph.

Additional to the question answering type of application like mentioned above, active data annotations can enable conversational agents to complete tasks like booking a room without having coupled APIs \cite{Simsek2018}. We demonstrate how a dialogue system can process the semantic API descriptions to guide a dialogue to book a room through an annotated IBE in the next use case.

\subsection{Active Data Consumption}
As a show case for the use of active data we built an API layer on top of the knowledge graph\footnote{\url{https://actions.semantify.it/api-docs/}}. This layer defines an entry point in form of a schema:SerachAction which points to the search API we provide. The response is a list of schema:Offers where the concrete manifestation depends on the type of search request. Every offer has another schema:Action attached which points to the action providers IBE. Trough that our graph acts like a broker between a user and the action provider where the business action is executed on the provider's side. Together with the action wrapper mentioned above, this allows the concept of "automatic direct booking" of hotel offers, which was published in \cite{kaerle2017annotation}. The hotel data in the graph, together with the annotated booking APIs of Easybooking and Feratel becomes directly bookable trough the annotations pointing towards the hotel's own booking API.

\subsection{Data Analysis}
\begin{figure}
\includegraphics[width=\linewidth]{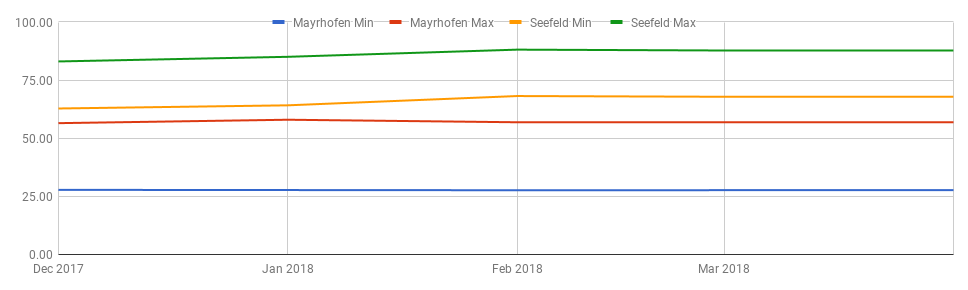}
\caption{Average minimum and maximum accommodation prices per person per night in Mayrhofen and Seefeld}
\label{fig:analysis}
\end{figure}

We store historical data in our knowledge graph as named graphs. This allows us to apply insightful analytics on the data. An example analysis is shown in Figure \ref{fig:analysis}. We analyzed the changes in average minimum and maximum accommodation prices per person per night in the regions of Mayrhofen and Seefeld between December 2017 and April 2018, based on the offer annotations we collected from the DMOs. The analysis shows that the prices in general are higher in Seefeld than Mayrhofen. There are no significant price fluctuations in Mayrhofen between different months. In Seefeld accommodation prices are visibly higher in February in comparison to other months.

\section{Conclusion \& Outlook}
\label{sec:Conclusion}
In this paper, we made a survey of our recent effort for building the Tirol Tourism Knowledge Graph from scratch. Starting with the definitions of schema.org subsets, we show how we then mapped touristic data sources and web APIs to schema.org and we present the idea of a publication heuristics for dynamic data. Finally we show four use cases where data from the knowledge graph is used by two different types of dialog systems, an eCommerce application and for statistical analysis over a time series of price development.

There are several learnings that can be taken from the survey we presented here. As already mentioned, the quality of a knowledge graph is highly depended on the quality of the data it contains. But the collection of touristic data frequently comes with lots of erroneous annotations. Importing the data from different source requires redundancies to be handled, hence a lot of data preprocessing before using the data. What we also learned is that the redundant storage of historical data in subgraphs makes sense for analysis of developments in tourism. Besides that, we learned that there are loads of use cases for knowledge graphs in tourism, most of which were not implemented in the course of that work but look promising.

Limitations of our approach are for example our method of mapping data sources to schema.org. Every source needs a own wrapper to be defined which is hardly scalable. The crawling of annotated content on websites is still very error-prone. But we are currently working on a more reliable crawler which performs validation and preprocessing by the time annotations are found, to improve data quality.

More future developments will improve data consolidation techniques when writing to the graph and add more incoming data sources, like other tourism service providers and internet booking engines. Apart from that we will show more use cases for the graph's data in the fields of assistant systems, machine learning and advanced reasoning.

 \bibliographystyle{splncs04}
 \bibliography{bib}
\end{document}